\documentclass[aps,twocolumn,showpacs,flushleft,amsmath,amssymb,nofootinbib]{revtex4}
\usepackage{subfig}
\usepackage{graphicx}
\usepackage{dcolumn} 
\usepackage{bm}      
\usepackage[active]{srcltx} 
\usepackage{hyperref} 
\usepackage{amsmath}
\usepackage{amssymb}
\usepackage{stmaryrd}
\usepackage{amsfonts}
\usepackage{mathrsfs}
\usepackage{txfonts}

\newcommand{\Ref}[1]{(\ref{#1})}

\newcommand{\eqa}{\begin{eqnarray}}
\newcommand{\neqa}{\end{eqnarray}}
\newcommand{\be}{\begin{equation}}
\newcommand{\ee}{\end{equation}}
\newcommand{\half}{\frac{1}{2}}

\newcommand{\suq}{\mathrm{SU}_q(2)}

\def\be{\begin{eqnarray}}
\def\ee{\end{eqnarray}}

\newcommand{\ca}{\mathcal A}

\newcommand{\ck}{\mathcal K}

\newcommand{\cp}{\mathcal P}

\newcommand{\g}{\gamma}
\newcommand{\G}{\Gamma}

\newcommand{\eps}{\epsilon}

\newcommand{\sig}{\sigma}

\renewcommand{\l}{\lambda}
\renewcommand{\L }{\Lambda}
\renewcommand{\o}{\omega}
\renewcommand{\O}{\Omega}

\newcommand{\rmd}{\mathrm d}

\newcommand{\lt}{\left}
\newcommand{\rt}{\right}

\newcommand{\lag}{\left\langle}
\newcommand{\rag}{\right\rangle}

\begin{document}

\sloppy

\title{\bf Cosmological Constant in LQG Vertex Amplitude}

\author{Muxin Han}

\affiliation{Centre de Physique Th\'eorique%
 \footnote{Unit\'e mixte de recherche (UMR 6207) du CNRS et des Universit\'es de Provence (Aix-Marseille I), de la Meditarran\'ee (Aix-Marseille II) et du Sud (Toulon-Var); laboratoire affili\'e \`a la FRUMAM (FR 2291).}, CNRS-Luminy Case 907,  F-13288 Marseille, EU}


\begin{abstract}
\noindent
A new q-deformation of the Euclidean EPRL/FK vertex amplitude is proposed by using the evaluation of the Vassiliev invariant associated with a 4-simplex graph (related to two copies of quantum SU(2) group at different roots of unity) embedded in a 3-sphere. We show that the large-j asymptotics of the q-deformed vertex amplitude gives the Regge action with a cosmological constant. In the end we also discuss its relation with a Chern-Simons theory on the boundary of 4-simplex.
\end{abstract}


\maketitle




\section{Introduction:} 

The spinfoam formalism is currently understood as a covariant formulation of Loop Quantum Gravity (LQG) \cite{book,rev,sfrevs,EPRL,FK}. In LQG community, it was commonly conjectured that one should make a q-deformation of the spinfoam amplitude (with quantum group) in order to implement the cosmological constant term in the theory \cite{sfrevs,smolin,new}. Such a conjecture was suggested by the lesson from 3d gravity and 4d topological field theory. In 3d gravity, the Turaev-Viro model \cite{TV} is a deformation of the Ponzano-Regge model \cite{PR} by the quantum group $\suq$ ($q$ is a root of unity). The partition function of the Turaev-Viro model is finite, and its large spin asymptotics give the 3d Regge action with a positive cosmological constant \cite{MT}. In 4d, the Crane-Yetter model \cite{CY} is a deformation of 4d SU(2) BF theory (the Ooguri model \cite{Ooguri}) by $\suq$ ($q$ is a root of unity). The partition function of the Crane-Yetter model is finite and shown to be the partition function of 4d SU(2) BF theory with a cosmological constant \cite{freidel}.

For 4d quantum gravity, there are early pioneer works for q-deformed LQG \cite{smolin}. In the spinfoam formulation, there are several proposals to make q-deformed spinfoam models, which are hoped to give the cosmological constant term in the semiclassical limit \cite{BCq,EPRLq,EPRLq2,DH}. In this paper, we propose a new q-deformation of the Euclidean EPRL/FK vertex amplitude by using the evaluation of the Vassiliev invariant associated with a 4-simplex graph (relats to the quantum group $\text{SU}_{q^{+}}(2)\otimes\text{SU}_{q^{-}}(2)$ with $q^\pm$ roots of unity). We also show that the large-j asymptotics of the q-deformed vertex amplitude gives the Regge action with cosmological constant. This result can be considered as an evidence supporting that the q-deformation of spinfoam amplitude implements the cosmological constant term in the framework of covariant LQG.

\section{Heuristic deformation:} 

Before we come to the systematic q-deformation of the amplitude, we first present a heuristic deformation of EPRL/FK vertex amplitude, to give an idea for obtaining the cosmological constant term in the spinfoam vertex amplitude.

Given a 4-simplex $\sig$, we label by $a,b=1,\cdots,5$ the five tetrahedra on the boundary of the 4-simplex, and denote by the pair $(a,b)$ the triangle shared by two tetrahedra $a$ and $b$. We assume the Barbero-Immirzi parameter $0<\g<1$, the Euclidean EPRL/FK vertex amplitude can be written in coherent state representation \cite{semiclassical} ($\pm$ stands for the self-dual/anti-self-dual contribution):
\be
A_\sig(k_{ab},n_{ab}):=(-1)^\chi\int\prod_{a=1}^5\rmd g_a^\pm\prod_{a<b}P^\pm_{ab}(k_{ab},g^\pm_a,n_{ab})\label{A}
\ee
where $(-1)^\chi$ is a sign defined by the diagrammatic calculus of SU(2) spin-network, $P^\pm_{ab}$ is a coherent propagator
\be
P_{ab}^\pm:=\lag j^\pm_{ab},-n_{ab}|(g^\pm_a)^{-1}g^\pm_b|j^\pm_{ab},n_{ba}\rag\label{P}
\ee
$g_a$ ($a=1,\cdots,5$) are $2\times2$ SU(2) matrices, and $|j,n\rangle$ is a coherent state in the spin-$j$ representation of SU(2)\cite{Perelomov}.  Here $\{k_{ab},n_{ab}\}$ with $j^\pm_{ab}=\frac{1\pm\g}{2}k_{ab}$ and $n_{ab}\in S^2$ is a set of boundary data for a vertex amplitude. The vector $j_{ab}n_{ab}$ is an oriented area vector of the triangle $(a,b)$ viewed at the tetrahedron $a$. The coherent state representation of EPRL/FK vertex amplitude is the starting point for the asymptotic analysis, and it turns out also to be useful in the analysis of quantum group spinfoam vertex.

\begin{figure}[h]
\begin{center}
\includegraphics[width=7cm]{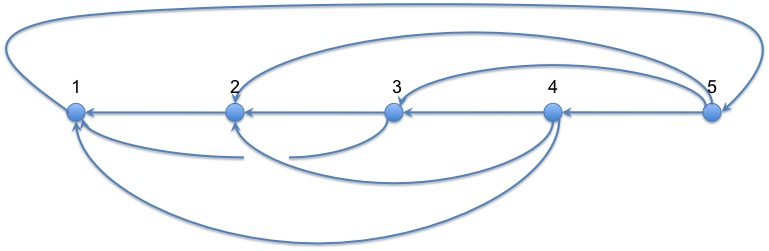}
\caption{The $\G_5^+$ graph with one crossing between $l_{31}$ and $l_{42}$.}
\label{gamma5}
\end{center}
\end{figure}

Now we make a heuristic modification of the EPRL/FK vertex amplitude: We consider a 4-simplex graph $\G_5^+$ (FIG.\ref{gamma5}). In FIG.\ref{gamma5} we order the 5 nodes on the paper from left to right, and connect the nodes by oriented links. A link oriented from the node $a$ to the node $b$ is denoted by $l_{ab}$. We notice that there is a crossing between the links $l_{31}$ and $l_{42}$, which motivate us to make the following modification of the coherent propagator $P_{31}^\pm$ and $P_{42}^{\pm}$. We define two operators $R^\pm$ on the SU(2) tensor representations $V_{j_{31}^\pm}\otimes V_{j_{42}^\pm}$ respectively\footnote{The original EPRL/FK amplitude doesn't know about the embedding of the 4-simplex spin-network, i.e. it doesn't depend on whether $l_{31}$ is over-crossing or under-crossing $l_{42}$. However the deformed amplitude does know about the embedding, thanks to the operators $R^{\pm}$.}:
\be
R^\pm:=\exp\lt[\frac{\pm4i}{(1\pm\g)^2}\o\sum_{j=1}^3 X_j^\pm\otimes X_j^\pm\rt]\label{R}
\ee
where $\o$ is a real dimensionless parameter (a deformation parameter), and $X_j^\pm$ are self-dual/anti-self-dual generators of Spin(4) with commutator $[X_j^\pm,X_k^\pm]=i\eps_{jkl}X_l^\pm$. We write formally that $R^{\pm}\equiv\sum_{R^\pm}R^\pm_{31}\otimes R^\pm_{42}$, and insert them by hand into the coherent propagators, i.e. we modify $P^\pm_{13}P_{24}^\pm$ by
\be
{}^\o\!P^\pm_{31}\ {}^\o\!P_{42}^\pm&=&\sum_{R^\pm}\lag j_{31}^\pm,-n_{31}|(g^\pm_3)^{-1}R^\pm_{31}g^\pm_1|j^\pm_{31},n_{13}\rag\nonumber\\
&&\times\ \lag j_{42}^\pm,-n_{42}|(g^\pm_4)^{-1}R^\pm_{42}g^\pm_2|j^\pm_{42},n_{24}\rag
\ee
while we leave the other coherent propagators unchanged as Eq.\Ref{P}. The modified (deformed) vertex amplitude is defined in the same way as Eq.\Ref{A} but with modified coherent propagator. We denote the modified vertex amplitude by $A^\o_\sig(k_{ab},n_{ab})$, which can written by
\be
A^\o_\sig(k_{ab},n_{ab})=\int\prod_{a=1}^5\rmd g_a^\pm\ \ck^\o_{31,42}\prod_{a<b}P^\pm_{ab}\label{VA0}
\ee
where $\ck_{31,42}^\o$ is a ratio
\be
\ck_{31,42}^\o=\frac{\prod_{\eps=\pm}{}^\o\!P^\eps_{31}\ {}^\o\!P_{42}^\eps}{\prod_{\eps=\pm}P^\eps_{13}P_{24}^\eps}.
\ee
We can then expand Eq.(\ref{R}) into power series of $\o$, which results in a power expansion of $\ck_{31,42}^\o$ in terms of the deformation parameter $\o$. A building block for constructing the power expansion of $\ck_{31,42}^\o$ is
\be
\prod_{ab=31;42}\frac{\lag j_{ab}^\pm,-n_{ab}|(g^\pm_a)^{-1}X^\pm_{k_1}\cdots X^\pm_{k_n}g^\pm_b|j^\pm_{ab},n_{ba}\rag}{\lag j_{ab}^\pm,-n_{ab}|(g^\pm_a)^{-1}g^\pm_b|j^\pm_{ab},n_{ba}\rag},\label{omegan}
\ee
which contributes the power expansion at the order $\o^n$.

By using the resolution of identity for coherent state $\dim(j)\int_{S^2}\rmd n|j,n\rangle\langle j,n|=1_j$, we can compute
\be
&&\frac{\lag j_{ab},-n_{ab}|g_a^{-1}X_{k_1}\cdots X_{k_n}g_b|j_{ab},n_{ba}\rag}{\lag j_{ab},-n_{ab}|g_a^{-1}g_b|j_{ab},n_{ba}\rag}\nonumber\\
&=&\frac{\dim(j_{ab})^{n-1}}{\lag j_{ab},-n_{ab}|g_a^{-1}g_b|j_{ab},n_{ba}\rag}\int_{(S^2)^n}\rmd n_1\cdots\rmd n_n\nonumber\\
&&\times \ \exp\Big[2j_{ab}\lt(\ln\langle -n_{ab}|g_a^{-1}|n_1\rangle+\cdots+\ln\lag n_{n}|g_b|n_{ba}\rag\rt)\Big]\nonumber\\
&&\times\ j_{ab}\frac{\lag -n_{ab}|g_a^{-1}\sig_{k_1}|n_1\rag}{\lag -n_{ab}|g_a^{-1}|n_1\rag}\cdots
j_{ab}\frac{\lag n_{n}| \sig_{k_n}g_b|n_{ba}\rag}{\lag n_{n}|g_b|n_{ba}\rag}\label{jn}
\ee
where we have used the following identity:
\be
&&\lag j,-n_1|g_1^{-1}X_kg_2|j,n_{2}\rag\nonumber\\
&=&j\lag-n_1|g_1^{-1}\sig_kg_2|n_{2}\rag\lag-n_1|g_1^{-1}g_2|n_{2}\rag^{2j-1}
\ee
We scale the spin $j_{ab}\mapsto \l j_{ab}$ and study the large-j asymptotic behavior of the integral in Eq.(\ref{jn}) as $\l\to\infty$. The leading asymptotics is determined by the critical point of the action
\be
S_0=2j_{ab}\lt[\ln\lag -n_{ab}|g_a^{-1}|n_1\rag+\cdots+\ln\lag n_{n}|g_b|n_{ba}\rag\rt]
\ee
The condition $\mathrm{Re}S_0=0$ gives the critical equations
\be
-g_an_{ab}=n_1=n_2=\cdots=n_n=g_bn_{ba}.\label{criticalS0}
\ee
The variations of the action $\delta S_0/\delta n_k$ vanishes automatically, once the above critical equations are satisfied. The asymptotics of Eq.(\ref{jn}) is given by the integrand evaluated at the critical point (critical equations). By using the following relation
\be
\frac{\lag -n_{ab}|g_a^{-1}\vec{\sig} g_b|n_{ba}\rag}{\lag -n_{ab}|g_a^{-1}g_b|n_{ba}\rag}=\frac{\tilde{n}_{ba}-\tilde{n}_{ab}+i\tilde{n}_{ab}\times \tilde{n}_{ba}}{1-\tilde{n}_{ab}\cdot \tilde{n}_{ba}}
\ee
where $\tilde{n}_{ab}=g_a n_{ab}$, we obtain the following asymptotic formula:
\be
&&\frac{\lag \l j_{ab},-n_{ab}|g_a^{-1}X_{k_1}\cdots X_{k_n}g_b|\l j_{ab},n_{ba}\rag}{\lag \l j_{ab},-n_{ab}|g_a^{-1}g_b|\l j_{ab},n_{ba}\rag}\nonumber\\
&\sim& \l j_{ab}(g_b{n}_{ba})^{k_1}\cdots\l j_{ab}(g_b{n}_{ba})^{k_n}[1+o(1/\l)]\label{replace}
\ee
Since Eq.(\ref{omegan}) is a product of two factors with $ab=31$ and $ab=42$, the building block in Eq.(\ref{omegan}) scales as $\l^{2n}$ as its leading large-j asymptotics. Moreover Eq.(\ref{omegan}) contributes the expansion at the order $\o^n$, thus [$\o^n\times$Eq(\ref{omegan})] doesn't scale asymptotically if we propose a scaling of $\o$ by $\o\mapsto\o/\l^2$.

From Eq.(\ref{replace}) we see that the asymptotic formula of a coherent state expectation value for $X_{k_1}\cdots X_{k_n}$ is given by simply replacing each $\vec{X}$ by $\l j_{ab}(g_b\vec{n}_{ba})$. Then we find that under the scaling $j_{ab}\mapsto\l j_{ab}$ and $\o\mapsto\o/\l^2$, the asymptotic formula for $\ck_{31,42}^\o$ as $\l\to\infty$ is obtained by considering the product $R^+R^-$ and replacing each $\vec{X}^\pm$ in $R^+R^-$ by $\l j_{ab}^\pm\vec{n}^\pm_{ba}$ ($\vec{n}^\pm_{ba}=g^\pm_{b}\vec{n}_{ba}$):
\be
\ck_{31,42}^\o&\sim&e^{i\o V_{31,42}}\lt[1+o(1/\l)\rt]\label{asymK}
\ee
where we denote
\be
V_{31,42}:=k_{31}\vec{n}^+_{13}\cdot k_{42}\vec{n}^+_{24}- k_{31}\vec{n}^-_{13}\cdot k_{42}\vec{n}^-_{24}.
\ee
recall that $j^\pm_{ab}=\frac{1\pm\g}{2}k_{ab}$. 

We write $\prod_{a<b}P^\pm_{ab}=e^S$ in the deformed vertex amplitude $A_\sig^\o$ in Eq.\Ref{VA0}, 
where $S$ is a ``spinfoam action'' used in the spinfoam asymptotic analysis \cite{semiclassical}
\be
S=\sum_{a<b}\sum_{\eps=\pm}2j_{ab}^\eps\log\lag -n_{ab}|(g^\eps_a)^{-1}g^\eps_b|n_{ba}\rag
\ee
The spinfoam action $S$ doesn't depend on $\o$. Thus under the scaling $k\mapsto\l k$, $\o\mapsto \o/\l^2$ and $\l\to\infty$, the $e^S$ part of the integrand is affected only by the scaling of the spins $k_{ab}$. The critical point of the action $S$ under $\l\to\infty$ is analyzed in \cite{semiclassical}. The critical equations from $S$
\be
\sum_b k_{ab}n_{ab}=0,\ \ \ \ g^\pm_a n_{ab}=-g^\pm_b n_{ba} \label{criticalS}
\ee
imply that (i) the closure of each tetrahedron and (ii) two neighboring tetrahedron are glued with each other at a triangle. Note that the critical equations Eq.\Ref{criticalS} from $S$ are consistent with the critical equations Eq.\Ref{criticalS0} from $S_0$. Suppose we fix a set of boundary data $\{k_{ab},n_{ab}\}$ corresponding to a non-degenerated flat 4-simplex Regge geometry, and also fix the dihedral angles between each pairs of neighboring tetrahedra (e.g. via imposing boundary state \cite{holo}), then there is a unique solution $(g^+_a,g^-_a)$ for the above critical equations. The solution specifies uniquely a bivector geometry of the 4-simplex up to an inversion. The bivector (at the center of 4-simplex) for each triangle $(a,b)$ is given by
\be
B_{ab}(\sig)=(B_{ab}^+,B^-_{ab})=\pm k_{ab}(g_a^+,g^-_a)(n_{ab},n_{ab})
\ee
Then one can see immediately the above $V_{31,42}$ evaluated at the critical point $(g_a^+,g_a^-)$ gives precisely the 4-volume of the 4-simplex $\sig$ (up to an overall constant)
\be
V_{31,42}\big|_{\text{critical}}=B^+_{31}\cdot B_{42}^+-B^-_{31}\cdot B_{42}^-=V_\sig
\ee
For a geometrical 4-simplex, this expression of 4-volume doesn't depend on the choice of triangle $(3,1)$ and $(4,2)$.

The asymptotics of the deformed vertex amplitude $A_\sig^\o$ is given by its integrand $\ck_{31,42}^\o e^S$ evaluated at the critical point satisfying both Eqs.(\ref{criticalS0}) and (\ref{criticalS}) from both actions $S$ and $S_0$. We have seen that the two critical equations Eqs.(\ref{criticalS0}) and (\ref{criticalS}) are consistent with each other. The action $S$ evaluated at the critical point gives the 4-simplex Regge action $iS_{\text{Regge}}=i\ell_p^2\sum_{a<b}\g k_{ab}\Theta_{ab}$ without cosmological constant. Eq\Ref{asymK} gives the asymptotic behavior of $\ck^\o_{31,42}$. Therefore we have the following large-j asymptotics
\be
A^\o_\sig\sim(\frac{2\pi}{\l})^{\frac{D}{2}}\frac{e^{\mathrm{ind} H}}{\sqrt{|\det H|}}e^{i\l\sum_{a<b}\g k_{ab}\Theta_{ab}}e^{i\o V_\sig}\lt[1+o(1/\l)\rt]\label{asymp1}
\ee
under $k_{ab}\mapsto\l k_{ab}$, $\o\mapsto \o/\l^2$ and $\l\to\infty$, where $H$ is the Hessian matrix of the spinfoam action $S$ and $D$ is the dimension of the integral. The above asymptotic formula manifests that the deformation parameter $\o$ is proportional to the cosmological constant $\L$ in Regge gravity. Note that the above Regge action with $\L$
\be
S_{\text{Regge},\L}=\ell_p^2\sum_{a<b}\g k_{ab}\Theta_{ab}+\L V_\sig
\ee 
corresponds to the Regge calculus approximation of continuous curved geometry with flat 4-simplices.

We now discuss the physical meaning of the scaling $k_{ab}\mapsto\l k_{ab}$, $\o\mapsto \o/\l^2$ and $\l\to\infty$, which leads us to the asymptotic formula Eq.\Ref{asymp1}. Given a cosmological constant $\L=1/\ell_c^2$ where $\ell_c$ is the cosmological length, the dimensionless parameter $\o$ has to be interpreted as $\o=\L\ell_p^2=\ell_p^2/\ell_c^2$ from the asymptotic formula Eq.(\ref{asymp1}). The spins $k_{ab}$ relate to the area $A_{ab}$ of the triangle shared by tetrahedra $a$ and $b$ by the relation $\g k_{ab}=A_{ab}/\ell_p^2$. Then the scaling $k_{ab}\mapsto\l k_{ab}$ can be understood as a scaling of the Planck length by $\ell_p^2\mapsto\l^{-1}\ell_p^2$ while keeping the area $A_{ab}$ fixed. The other scaling $\o\mapsto \o/\l^2$ combined with $\ell_p^2\mapsto\l^{-1}\ell_p^2$ results in the scaling of the cosmological length $\ell_c^2\mapsto\l\ell_c^2$. As $\l\to\infty$, we see that the asymptotic formula Eq.\Ref{asymp1} is valid in the regime where the area $A_{ab}$ is much larger than the Planck area $\ell_p^2$ but much smaller than the cosmological area $\ell_c^2$. The assumption that the cosmological length $\ell_c$ is much larger than the physical scale of the 4-simplex is the reason why we can approximate the local geometry with a flat 4-simplex given by the critical equations Eq.\Ref{criticalS} and the boundary data $\{k_{ab},n_{ab}\}$.

\section{q-deformation and Vassiliev invariants: }

From the above derivation, we have seen that the expected cosmological constant term comes from the insertion of the operator $R^\pm$ in the vertex amplitude, which is responsible for the crossing in the spin-network graph $\G_5^+$. Here we present a more systematic deformation of the EPRL/FK vertex amplitude by using the evaluation of Vassiliev invariants \cite{vassiliev} (see also \cite{freidel} for a brief introduction). The resulting q-deformed vertex amplitude gives has the same asymptotic behavior as the above heuristic deformation.

Let's recall Eq.\Ref{A} and carry out the integration over $g^\pm_a$, we obtain
\be
A_\sig(k_{ab},n_{ab})=\sum_{\{i_a^\pm\}}\{15j\}_{i^\pm_a}^\pm\prod_{a=1}^5f_{i_a^\pm}(j^\pm_{ab},n_{ab})\label{A1}
\ee
where $\{15j\}_{i^\pm_a}^\pm$ denotes two copies of SU(2) 15j symbol with spins $j^\pm_{ab}$ and intertwiners $i^\pm_a$, and $f_{i_a^\pm}(j^\pm_{ab},n_{ab})$ denotes two copies of SU(2) intertwiner $i^\pm_a$ in the coherent state representation. 

We define a deformation of the vertex amplitude by simply replace the 15j symbols in Eq.(\ref{A1}) by two q-deformed 15j symbols with $q^\pm$ at \emph{different} roots of unity. Therefore we define the $q$-deformed EPRL/FK vertex amplitude by
\be
A^q_\sig(k_{ab},n_{ab}):=\sum_{\{i_a^\pm\}}\{15j\}_{i^\pm_a,q^{\pm}}^\pm\prod_{a=1}^5f_{i_a^\pm}(j^\pm_{ab},n_{ab})\label{Aq}
\ee
The q-deformed 15j symbols are obtained from the evaluation of a 4-simplex spin-network with the corresponding Vassiliev invariant. Here we briefly describe the procedure for the construction.

Let $X$ be a 1-dimensional oriented compact manifold (an oriented graph). A chord diagram with support $X$ is defined by the union $C=D\cup X$, where $D$ (dash lines) is a (non-planar) graph with end points on $X$, and the graph $D$ has only univalent and trivalent vertices. The degree of the chord diagram $C$ is defined by the half of the number of vertices in $D$. We define a vector space $\ca_n(X)$ generated by all the chord diagrams with degree $n$, subject to the some relations \cite{freidel,vassiliev,natan}

\begin{figure}[h]
\begin{center}
\includegraphics[width=5cm]{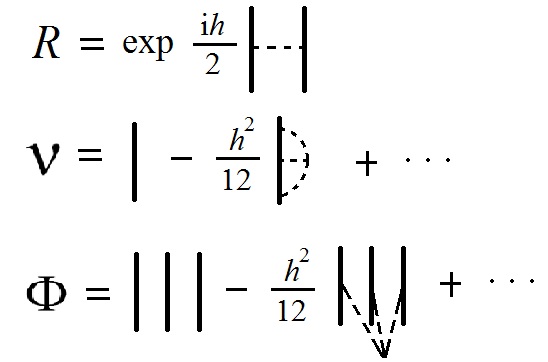}
\caption{The building blocks for Vassiliev invariant.}
\label{block}
\end{center}
\end{figure}

The space of chord diagrams is used to define the universal Vassiliev invariant for the framed links. Given a deformation parameter $q=e^{ih}$, the Vassiliev invariant $Z$ assigns to any framed link $X$ a formal power series $Z(X)=\sum_{n=0}^\infty h^n Z_n(X)$, where the coefficients $Z_n(X)\in \ca_n(X)$ is a linear combination of degree-$n$ chord diagrams. Given the link $X$, we need three types of building blocks to construct $Z_n(X)$ to each order: (1) For each crossing in $X$ we assign a braiding $R\in \cp_2$; (2) For each maximum or minimum in $X$ we assign a unknot $\nu^{-\half}\in\cp_1$; (3) There is also an associator $\Phi\in\cp_3$ \cite{Drinfeld}. Here $\cp_n$ denotes the space of the series of chord diagrams based on $n$ lines in $X$. These building blocks are expressed as power series FIG.\ref{block} (In the exponential for $R$-matrix, the product of two chord diagrams is defined by placing one diagram on top of the other.)

\begin{figure}[h]
\begin{center}
\includegraphics[width=5cm]{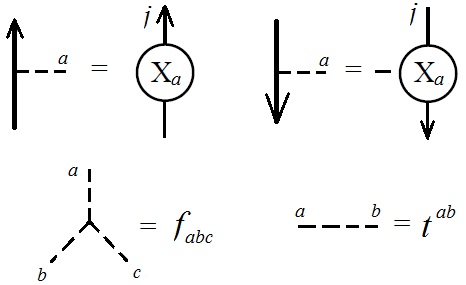}
\caption{Evaluation of Vassiliev invariant.}
\label{evaluation}
\end{center}
\end{figure}

Given a compact Lie group $G$ and a spin-network $s$ based on the oriented graph $X$, for each chord diagram based on $X$, we can define the evaluation map $\O_{G,s}$ given by FIG.\ref{evaluation}. Here $X_a$ is a basis of the Lie algebra $Lie(G)$ with structure constant $f_{abc}$, and $t^{ab}X_aX_b$ is the quadratic casimir of $Lie(G)$. It turns out that the evaluation $\O_{G,s}$ of links gives the same result as the Reshetikhin-Turaev evaluation of the link associated with the quantum group $U_q(G)$ \cite{vassiliev,AF,RT}.

\begin{figure}[h]
\begin{center}
\includegraphics[width=8cm]{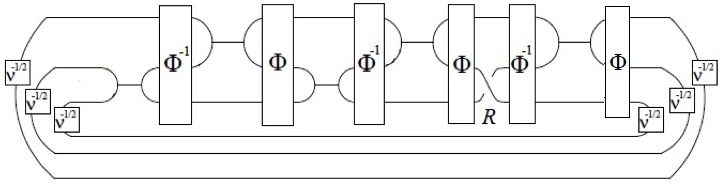}
\caption{The evaluation of 4-simplex graph via Vassiliev invariant.}
\label{4-simplex}
\end{center}
\end{figure}

For a 4-simplex SU(2) spin-network based on the graph $\G_5^+$, the correspond 15j symbol $\{15j\}_{i_a,q}$ is given by the evaluation of FIG.\ref{4-simplex} with appropriate insertions of $R$-matrix, associators $\Phi$, and unknots $\nu^{\half}$. We evaluate FIG.\ref{4-simplex} for both self-dual and anti-self-dual sector, and insert them in the definition of the q-deformed vertex amplitude Eq.(\ref{Aq}). As we did for the heuristic deformation $A_\sig^\o$, we expand the q-deformed vertex amplitude $A^q_\sig$ into a power series of $\o$. For the braiding $R$-matrix responsible for the only crossing in FIG.\ref{4-simplex}, its evaluation coincides with Eq.(\ref{R}) used in the heuristic deformation, if we choose the deformation parameter $q^\pm=e^{ih^{\pm}}$ such that
\be
h^{\pm}=\pm\frac{8}{(1\pm\g)^2}\o.\label{h}
\ee

In the following, we show that both the associator $\Phi$ and unknot $\nu$ don't contribute the leading asymptotic behavior of $A_\sig^q$ under the scaling $k_{ab}\mapsto\l k_{ab}$, $\o\mapsto \o/\l^2$ and $\l\to\infty$. First of all, the SU(2) evaluation of unknot $\nu$ can be expanded as a power series of $h$ by (see e.g. \cite{vassiliev})
\be
\nu=\sum_{n=0}^\infty q_n(c)h^{2n}\label{nu}
\ee 
where $c$ is the quadratic casimir of su(2). the polynomial function $q_n$ relates to the Bernoulli polynomial $B_{2n+1}$ by
\be
q_n\lt(\frac{x^2-1}{2}\rt)=\frac{2}{(2n+1)!}\frac{B_{2n+1}\lt[\half x+\half\rt]}{x}
\ee
In the scaling of spins $k_{ab}\mapsto\l k_{ab}$, the quadratic casimir scales as $\l^2$. Then $q_n(c)$ scales as $\l^{2n}$ since $B_{2n+1}[\l x]\sim \l^{2n+1}B_{2n+1}[x]$ as $\l\to\infty$. As a result each term $q_n(c)h^{2n}$ in Eq.\Ref{nu} scales as $\l^{-2n}$ by taking into account the scaling $\o\mapsto \o/\l^2$. Thus the leading asymptotic behavior of $A_\sig^q$ only sees $\nu=1$ since all the higher order corrections only contribute $o(1/\l)$-terms in Eq.\Ref{asymp1} as $\l\to\infty$.

The perturbative expansion of the associator $\Phi$ can be presented in terms of chord diagrams in FIG.\ref{block}, where the degree-$n$ chord diagram at each $h^n$-order is build by connecting the 3-valent vertices of dashed lines in FIG.\ref{evaluation}. There are $2n$ vertices in each degree-$n$ diagram, in which there are $m$ vertices are attached to the framed links. Thus $2n-m$ is the number of internal 3-valent vertices and $2n-m>0$ for a nontrivial chord diagram. When we scale of spins $k_{ab}\mapsto\l k_{ab}$ and $\l\to\infty$, the evaluation of each vertex attached to a framed link gives a factor of $\l j^\pm_{ab} \vec{n}^\pm_{ab}$ as its leading asymptotics, since on each link the su(2) generator $X_a$ is sandwiched by SU(2) coherent states. Thus for each degree-$n$ diagram in the perturbative expansion of $\Phi$, the scaling of spins $k_{ab}\mapsto\l k_{ab}$ leads to a scaling $\l^{m}$ of the diagram, while the other scaling $\o\mapsto \o/\l^2$ contributes $h^n\mapsto \l^{-2n}h^n$. Thus the overall scaling of each term is $\l^{-(2n-m)}$, from which we see that the nontrivial diagrams in $\Phi$ only contributes to the $o(1/\l)$-terms in the asymptotic formula as $\l\to\infty$.

The above power-counting shows that we can take $\Phi=1$ and $\nu=1$ for the asymptotic analysis of the q-deformed vertex amplitude $A^q_\sig$. By the coincidence of the R-matrix between $A^q_\sig$ and $A^\o_\sig$, the asymptotic analysis of $A^q_\sig$ reduces to the previous analysis of heuristic deformation $A^\o_\sig$, i.e. under the scaling $k_{ab}\mapsto\l k_{ab}$, $\o\mapsto \o/\l^2$ and $\l\to\infty$, $A_\sig^q$ and $A^\o_\sig$ have the same asymptotic behavior. Thus we can write down the asymptotic formula of the q-deformed vertex amplitude with a given Regge boundary data:
\be
A^q_\sig\sim(\frac{2\pi}{\l})^{\frac{D}{2}}\frac{e^{\mathrm{ind} H}}{\sqrt{|\det H|}}e^{i\l\sum_{a<b}\g k_{ab}\Theta_{ab}}e^{i\o V_\sig}\lt[1+o(1/\l)\rt].
\ee

\begin{figure}[h]
\begin{center}
\includegraphics[width=7cm]{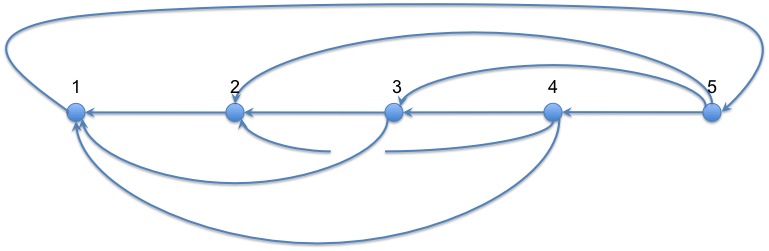}
\caption{}
\label{gamma5-}
\end{center}
\end{figure}

Before conclusion, we would like to point out an interesting fact: there is another possibility to obtain the same asymptotics from another q-deformation. We use the deformation parameter $h^{\pm}=\frac{8}{(1\pm\g)^2}\o$ instead of Eq.\Ref{h}, but evaluate the the self-dual and anti-self-dual 15j symbols on different graphs, i.e. we evaluate the self-dual sector on the $\G_5^+$ graph as before but evaluate the anti-self-dual sector on the $\G_5^-$ graph FIG.\ref{gamma5-} with the opposite crossing (with braiding $R^{-1}$) to the one in $\G_5^+$. Then it is not hard to see that the resulting q-deformed vertex amplitude has the same asymptotic behavior as the above up to higher order in $\l^{-1}$.

\section{Conclusion and Discussion:} 

To summarize, in this paper we propose a new q-deformation of the Euclidean EPRL/FK spinfoam vertex amplitude. The concrete construction uses the evaluation of the Vassiliev invariant from 4-simplex graph. We also show that the asymptotics of the q-deformed vertex amplitude gives the Regge gravity with a cosmological constant (from Regge calculus using flat 4-simplices) in the regime that the physical scale of the 4-simplex is much greater than the Planck scale $\ell_p$ but much smaller than the cosmological area $\ell_c$.

The Vassiliev invariants of links come from the Feymann diagrams of perturbative Chern-Simons theory, for evaluating the link observables \cite{AF,natan}. The q-deformation of the 15j symbol employed above can be viewed as a Chern-Simons expectation value of a 4-simplex spin-network. Moreover we suppose the boundary of the 4-simplex under consideration is a 3-sphere $S^3$, then the q-deformed vertex amplitude for this 4-simplex is given by the following expectation value of a Chern-Simons theory (with gauge group $\text{Spin(4)}=\text{SU(2)}\times \text{SU(2)}$) on the boundary 3-manifold:
\be
A^q_\sig=\int\Psi[A^\pm]\ e^{\frac{2\pi i}{h^+}S_{CS}[A^+]+\frac{2\pi i}{h^-}S_{CS}[A^-]}DA^\pm
\ee
where $S_{CS}[A]$ is the SU(2) Chern-Simons action, and $\Psi[A^\pm]$ is a projective spin-network functions on Spin(4) holonomies \cite{DL} associated with a 4-simplex graph $\G_5^+$ (or two graphs $\G_5^\pm$) imbedded in the boundary 3-sphere. Interestingly, this result also relates to an old idea by L. Smolin et al (see \cite{smolin}).

In addition, although all the discuss in this paper concerns only a single 4-simplex, the asymptotic analysis can be done also for a triangulation with arbitrary many 4-simplices, which results in a Regge action with a cosmological constant (from the Regge calculus with flat simplices) on the triangulation. The detailed analysis will be reported in \cite{toappear}.

Finally we note that the scaling $k_{ab}\mapsto\l k_{ab}$, $\o\mapsto \o/\l^2$ used in this paper leads us to the Regge calculus with flat 4-simplex, which is an approximation of curved geometry in presence of a cosmological constant. It would be interesting to find the relation between the q-deformed vertex amplitude and a curved 4-simplex with constant curvature, in analogy with the 3d case (see e.g. \cite{woodward}). We leave this point to the future research.

\section*{Acknowledgments}

The author would like to thank W. Wieland for sharing his idea \cite{wolfgang} about Chern-Simons theory, and thank L. Freidel for his comment on the early version of this paper. The author would also like to thank E. Bianchi, R. Coquereaux, K. Noui, A. Perez, S. Speziale, C. Rovelli, and M. Zhang for discussions and communications.

\end{document}